\def\figsize{9.5cm}

\documentclass[twocolumn,amsmath,nofootinbib]{revtex4} 
\usepackage{graphicx}
\usepackage{dcolumn}
\usepackage{bm}

\newcommand{\EQ}{\begin{equation}}
\newcommand{\eq}{\end{equation}}
\newcommand{\EQA}{\begin{eqnarray}}
\newcommand{\eqa}{\end{eqnarray}}

\newcommand{\AR}{\renewcommand {\arraystretch}{1.5}
\begin{array}{l}}
\newcommand{\bAR}{\renewcommand {\arraystretch}{2}
\begin{array}{l}}
\newcommand{\ARc}{\renewcommand {\arraystretch}{1.5}
\begin{array}{c}}
\newcommand{\bARc}{\renewcommand {\arraystretch}{2}
\begin{array}{c}}
\newcommand{\ar}{\end{array} \renewcommand {\arraystretch}{1}}

\begin{document}
\input{epsf.sty}

\vspace*{.15in}
\noindent
PHYSICAL REVIEW LETTERS
\vskip .20in
\noindent


 \title{Negative Energies and a Constantly Accelerating Flat Universe}

\author{Henry-Couannier, Tilquin, Ealet, Tao }

\address{CPPM, C.N.R.S. Université de la méditerranée - Luminy,case 907, Marseille, France}


\begin{abstract}



It has been shown that in the context of General Relativity (GR) enriched with a new set of 
discrete symmetry reversal conjugate metrics,  negative energy states can be rehabilitated while avoiding 
the well-known instability issues. We review here some cosmological implications of the model and 
confront them with the supernovae and CMB data. The predicted flat universe constantly accelerated expansion
phase is found to be in rather good agreement with the most recent cosmological data.

\end{abstract}

\maketitle







   

\section{Introduction}

One of the most challenging tasks in contemporary physics is to understand the 
observational results indicating that we are living in a flat accelerating
 universe. The most popular interpretation is that we are dominated by a homogeneous
component with negative pressure often called dark energy.
The supernovae data from \cite{sn} indicate that the equation of state of this dark energy 
is compatible with the 'concordance model' 
($\Omega_M = 0.3$, $\Omega_{\Lambda}$ = 0.7, w=p/$\rho$=-1) but a careful
interpretation of the data \cite{nous} shows that a component with w $<$ -1
is allowed. Models with very exotic $w(z)$ may come from modified gravity and have very
different consequences for the fate of the Universe\cite{LinderG}. 
Such models are quite unsatisfactory since they inevitably lead to a generic violation of  positive energy
conditions resulting in vacuum quantum instabilities \cite{Lin1} \cite{Cald} \cite{Fram} \cite{Carr}.

However, alternative proposals have been made  \cite{fhc2} \cite{Lin2} where 
the local instability issue can be avoided because the interaction between the 
positive and negative energy universes is global.
The model presented in \cite{fhc2} is particularly attractive since the conjugate universe and its negative energy
content are not introduced by hand but emerge as a result of imposing new symmetries on 
the initial action.  It is also very predictive: flatness is mandatory and a constant acceleration
phase is one of the very few mathematical possibilities in such a tightly constrained theoretical framework.   
 We confront in this paper its predictions to present observationnal data.

\section { Motivations for a modified GR} 
Investigation of negative energies in Relativistic 
Quantum Field Theory (QFT) indicates that the correct theoretical framework should be found in a modification of
 General Relativity (GR) \cite{fred}.

- TheoreticaI motivations

In second quantification, all relativistic field equations admit genuine negative energy field solutions 
creating and annihilating negative energy quanta. Unitary time reversal links these fields to the positive energy ones.
 The unitary choice, usual for all other symmetries in physics, also allows us to avoid the well known paradoxes
 associated with time reversal.
Positive and negative energy fields vacuum divergences encountered after second quantization, are 
unsurprisingly found to be exactly opposite. The negative energy fields action must be maximised.
However, there is no way to reach a coherent theory involving negative energies in flat-spacetime. Indeed,
if positive and negative energy scalar fields are time reversal conjugate, so must be their Hamiltonian densities and actions.
 This is only possible in the context of GR thanks to the 
metric transformation under discrete symmetries.

- Phenomenological motivations

In a mirror negative energy world, whose fields remain non coupled to our world positive energy fields, stability is insured
 and the behavior of matter and radiation is as usual. Hence, it is just a matter of convention to define each one as a positive
 or negative energy world. Otherwise, if they interact gravitationally, promising phenomenology is expected. Indeed, many outstanding enigmas
indicate that repelling gravity might play an important role 
in physics: flat galactic rotation curves, the Pioneer effect, 
the flatness of the universe, acceleration and its voids, etc... But negative energy states never manifested themselves up to now, suggesting
 that a barrier is at work preventing the two worlds to interact except through gravity.

- A modified GR to circumvent the main issues

A trivial cancellation between vacuum divergences is not acceptable since the Casimir effect shows
 evidence for vacuum fluctuations. But the positive and negative energy worlds could be maximally gravitationally coupled in such a way as 
to produce at least exact cancellations of vacuum energies gravitational effects.
 Also, a generic catastrophic instability issue arises whenever quantum positive and negative energy fields 
are allowed to interact. If we restrict the stability issue to the modified gravity, this disastrous scenario is avoided. 
Finally, allowing both positive and negative energy virtual photons to propagate the electromagnetic interaction, simply makes it disappear. 
The local gravitational interaction is treated very differently in our modified GR .So this unpleasant feature is also avoided.

\section{Conjugate Worlds Gravitational Coupling }

Ref.~\cite{Wein} shows that time reversal does not 
affect a scalar action. However, if the inertial coordinates $\xi ^\alpha $ 
are transformed in a non-trivial way: 

\begin{equation}
\xi ^\alpha \mbox{ }\mathop \to \limits^T \tilde {\xi }_T ^\alpha 
\label{eq0}
\end{equation}

where $\tilde {\xi }^\alpha \ne \xi_\alpha $, 
metric terms are affected and the action is not expected 
to be invariant under $T$. Having two conjugate inertial coordinate 
systems, two time reversal conjugate metric 
tensors can be built: 

\begin{equation}
g_{\mu \nu } =\eta _{\alpha \beta } \frac{\partial \xi ^\alpha }{\partial 
x^\mu }\frac{\partial \xi ^\beta }{\partial x^\nu },\quad
\tilde {g}_{\mu \nu } =\eta _{\alpha \beta } \frac{\partial \tilde {\xi 
}^\alpha }{\partial x^\mu }\frac{\partial \tilde {\xi }^\beta }{\partial 
x^\nu }
\label{eq1}
\end{equation}

Then, a new set of fields couples to the new $\tilde {g}_{\mu \nu } $ 
metric field. The total action is the sum of $I_M $, the usual action 
for matter and radiation in 
the external gravitational field $g_{\mu \nu } $, $\tilde {I}_M $ the action 
for matter and radiation in the external gravitational field $\tilde {g}_{\mu \nu } $
 and the actions $I_G +\tilde {I}_G $ for the gravitational fields
alone. The conjugate actions are 
separately general coordinate scalars and adding the two pieces 
is necessary to obtain a discrete symmetry reversal invariant total 
action. $g_{\mu \nu } $ and $\tilde {g}_{\mu \nu } $ are linked since these are symmetry reversal 
conjugate objects, explicitly built out of space-time coordinates. We 
postulate that
there exists a privileged general coordinate system such that $\tilde {g}_{\mu \nu } $
 identifies with $g^{\mu \nu }$, where for instance a discrete time reversal 
transformation applies as $x^0\to -x^0$.  In this system, varying the action, 
applying the extremum 
action principle and making use of the relation 
$\delta g^{\rho \kappa }\left( x \right)=-g^{\rho \mu }\left( x \right)g^{\nu \kappa }\left( x \right)\delta g_{\mu \nu } \left( x \right)$
 leads to a modified
Einstein equation only valid in the privileged  coordinate 
system.
This equation is not general covariant and not intended to be so. 

 \section{Modified cosmology}

Following the method outlined in the previous section, local solutions satisfying the symmetry invariance
 requirements under Parity or space/time exchange transformations have been found and interpreted in  
~\cite{fhc2}. In the case of cosmology and time reversal, there exists one 
global privileged coordinate system where a couple of purely time dependent 
time reversal conjugate solutions can be derived from the couple of 
conjugate actions. The existence of a time reversal conjugate universe 
was also suggested a long time ago in Ref.~\cite{Sak}.
The only possible privileged coordinate system where both 
metrics are spatially homogeneous and isotropic is the flat Cartesian one:

\begin{equation}
d\tau ^2=B(t) dt^2-A(t) dx^2
\label{eq5}
\end{equation}

The privileged coordinate system is the conformal time 
system where $B=A$. Expressing in the polar coordinate system, the modified 
cosmological Einstein equations are:
\begin{equation}
3A\left( {-\frac{\ddot {A}}{A}+\frac{1}{2}\left( {\frac{\dot {A}}{A}} 
\right)^2} \right)-\frac{3}{A}\left( {\frac{\ddot {A}}{A}-\frac{3}{2}\left( 
{\frac{\dot {A}}{A}} \right)^2} \right)=0
\end{equation}

The purely time dependent scale factor evolution is then driven (a nondimensional time unit is used) 
by the 
following differential equations in the three particular domains:

\begin{equation}
 \mbox{ }a<<1\Rightarrow \ddot {a}\propto \frac{3}{2}\frac{\dot {a}^2}{a}\Rightarrow a\propto 1/t^2\mbox{ where }t<0 ,
\end{equation}

\begin{equation}
\mbox{ }a\approx 1\Rightarrow \ddot {a}\propto \frac{\dot {a}^2}{a}\Rightarrow a\propto e^t  ,
\end{equation}

\begin{equation}
 \mbox{ }a>>1\Rightarrow \ddot {a}\propto \frac{1}{2}\frac{\dot {a}^2}{a}\Rightarrow a\propto t^2\mbox{ where $t$ }>\mbox{ 0} \\ .
\end{equation}

We check that $t \to -t$ implies 
$1/t^2 \to  t^2$ but also $e^t\to e^{-t}$, as required. 
Flatness is the main prediction of this model.  A striking and very uncommon feature is that the evolution of the scale factor is completely independent
of the matter and radiation content in the two universes. 
 In particular, the observed flatness can no longer be translated into the usual estimation of $\Omega_m=1$
from the WMAP data \cite{WMAP}. 
According the $t^2$ evolution, we are 
most probably living in a constantly accelerating universe. Our and the conjugate universe
either have crossed in the past or will cross each other and time reversal will 
occur in the future.  
Our universe is accelerated without any need for a cosmological
constant or dark energy component 
No source enters our cosmological equation except at the crossing time where  a small perturbation is 
needed to start the non stationary cosmological solutions.

\section {Application to supernovae data}
We have confronted this model with the existing supernovae data published by \cite{sn}. The luminosity 
distance, for a flat constantly accelerated ($t^2$ evolution) universe is:
\begin{equation}
d_L = a_0 r_l (1+z) = \frac{2}{H_0}(\sqrt{1+z} - 1)(1+z) \\
\end{equation}

\begin{figure} [ht]
\vskip-0.5cm
\centerline{\epsfxsize=\figsize\epsffile{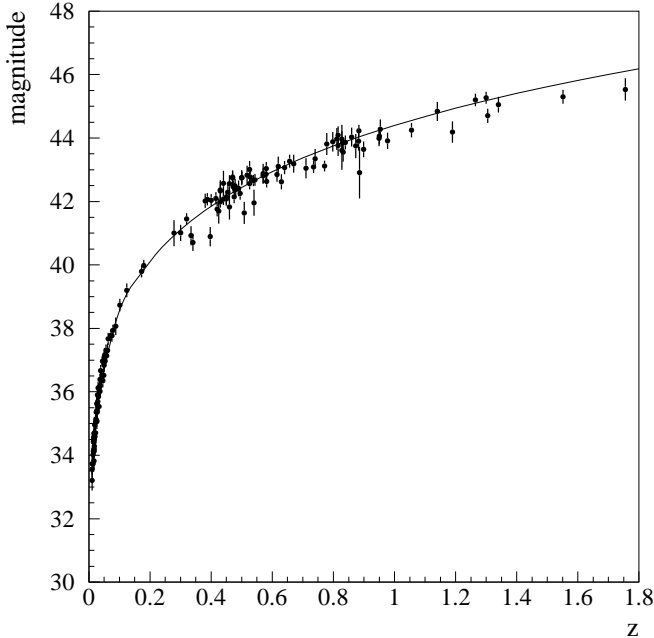}}
\vskip-0.5cm
\caption{\label{fig:result}\footnotesize%
The distance luminosity prediction of this model compared to the present supernovae data}
\end{figure}

Figure~\ref{fig:result} shows the fitted magnitude versus redshift curve from the model, 
where the only free parameter is the normalisation parameter $\mathrm{m_s}$, 
compared to the observational data derived from the gold sample of \cite{sn}. 
The quality of the fit is estimated with the computed $\mathrm{\chi^2}$ to be at a
$\mathrm{2\%}$ confidence level (CL). This 
should be compared to the $\mathrm{11\%}$ confidence level found by
using the standard $\mathrm{\Lambda CDM}$ model with
no prior. Fitting our model on the SCP data \cite{scp} leads to a $\mathrm{59\%}$
confidence level compatibility, to be compared with a $\mathrm{56\%}$
confidence level using the  $\mathrm{\Lambda CDM}$ model.

To go further in this analysis, we parametrise the scale factor
evolution as a simple power low $\mathrm{a\propto t^\alpha}$. The luminosity
distance reads:
\begin{equation}
d_L = \frac{1}{H_0}\frac{\alpha}{\alpha-1}((1+z)^{1-1/\alpha}- 1) (1+z)\\
\end{equation}
Fitting simultaniously $\mathrm{m_s}$ and $\alpha$ on the Riess gold data sample
gives $\mathrm{\alpha=1.41\pm0.13}$ which is $\mathrm{4.5\sigma}$ away from
the predicted value. Assuming a simple variation of
the intrinsic supernovae magnitude versus the redshift $\mathrm{\partial m_s / \partial z = 0.15}$
which is below the current statistical error, leads to a systematical error :
 $+0.5-0.3(syst.)$ for this model, both for Riess \cite{sn} and SCP \cite{scp} data.
This shows that the power law parametrisation is much more
sensitive to systematical errors than the standard fitting procedure including
evolution of the equation of state \cite{sn}. In conclusion, the current precision is
not sufficient to discriminate between the $\mathrm{\Lambda CDM}$ model and the model presented here.

We have then investigated the expected sensitivity with future SN projects.
We simulate the SNLS \cite{snls1} \cite{snls2} experiment expecting about 700 SNIa with redshifts up to about
1, adding 300 simulated nearby SNIa from the future SN factory project and using
$\mathrm{\Lambda CDM}$ as
fiducial model.  We get $\mathrm{\alpha=1.26\pm0.04(stat.)\pm 0.3(syst.)}$,
where the systematical error has been evaluated using a $10\%$ evolution on the
intrinsic magnitude with redshift. Thus the SNLS experiment will be able to
distinguish between the $\mathrm{\Lambda CDM}$  model and this one at a 3 sigma level.

The SNAP/JDEM \cite{SNAPweb} mission will observe about 2000 SNIa with redshifts up to 1.7. The
intrinsic evolution of SNIa magnitudes is expected to be controlled at the percent level.  
Again, using  $\mathrm{\Lambda CDM}$ as a fiducial model and 300 SNIa from
the SN factory model gives:$\mathrm{\alpha=1.24\pm0.02(stat.)\pm 0.04(syst.)}$, 
where the systematical error is coming from a $2\%$ remaining possible evolution of
the intrinsic magnitude. The SNAP/JDEM mission will thus definitely answer the question
of the compatibility of this new model with supernovae observations.

Finally, as explained in \cite{fhc2}, the constantly accelerated evolution could be affected by Pioneer like effets 
(as it is indeed in our neighborhood) resulting in locally inverted evolutions of space-space metric elements. 
This should result in a systematical drift proportional to the relative photons time of flight accross the regions with inverted 
cosmological regime along their path. The effect predominantly affects the higher redshift part of the path, when 
the matter structure (living in the inverted regime) occupied a relatively more important fraction of space, so a 
jerk behaviour is a natural outcome of the model. This could account for the difference between the constantly accelerated
regime prediction of this model and the current data favoring a recent transition from a decelerating to 
an accelerating universe.

\section{Conclusion}

Introducing discrete symmetries in the context of General Relativity not only allows to solve 
many long lasting theoretical issues such as negative energies and stability, QFT vacuum divergences and the 
cosmological constant but also leads to very remarkable phenomenological predictions.
A constantly accelerating necessarily flat universe is a natural outcome of the model
 and cannot be excluded by present data 
The large scale structure formation and evolution need to be completely revisited in the new context 
where the rules of the game are significantly modified due to the interactions between conjugate density fluctuations.
Last, as shown in \cite{fhc2}, the space/time exchange symmetry allows the derivation of a propagating solution 
and the clarification of  the status of tachyonic representations.
The published supernovae data are not in disagreement with this model.
The one is very sensitive to systematical effects.
 Therefore, only the SNAP/JDEM mission should be able to distinguish
without ambiguity between standard cosmology and a constantly accelerated regime as predicted by the new model
in the case where no pioneer-like theoretical systematical effect is to be expected.

\end{document}